\documentclass[12pt,preprint]{aastex}

\begin{document}

\title{Planetary systems around close binary stars: the case of the very 
dusty, Sun-like, spectroscopic binary BD+20~307}

\author{B. Zuckerman\altaffilmark{1,2}, Francis C.\ 
Fekel\altaffilmark{3}, Michael H.\ Williamson\altaffilmark{3}, Gregory 
W.\ Henry\altaffilmark{3}, M. P. Muno\altaffilmark{4}}

\altaffiltext{1}{Department of Physics \& Astronomy, University of 
California, Los Angeles CA 90095}
\altaffiltext{2}{UCLA Center for Astrobiology}
\altaffiltext{3}{Center of Excellence in Information Systems, Tennessee 
State University, 3500 John A.\ Merritt Blvd., Box 9501, Nashville, TN
37209}
\altaffiltext{4}{Space Radiation Laboratory, California Institute
of Technology, Pasadena, CA 91125-4700}

\begin{abstract}

Field star BD+20~307 is the dustiest known main sequence star, based on 
the fraction of its bolometric luminosity, $\sim$4\%, that is emitted at 
infrared wavelengths.  The particles that carry this large IR luminosity 
are unusually warm, comparable to the temperature of the zodiacal dust in 
the solar system, and their existence is likely to be a consequence of a 
fairly recent collision of large objects such as planets or planetary 
embryos.  Thus, the age of BD+20~307 is potentially of interest in 
constraining the 
era of terrestrial planet formation.  The present project was initiated 
with an attempt to derive this age using the 
{\it Chandra X-ray Observatory} to measure the X-ray flux of BD+20~307 in 
conjunction with extensive photometric and spectroscopic monitoring 
observations from Fairborn Observatory.  However, the recent realization 
that BD+20~307 is a short period, double-line, spectroscopic binary whose 
components have very different lithium abundances, vitiates standard 
methods of age determination.  We find the system to be metal-poor; this, 
combined with its measured lithium abundances, indicates that BD+20~307 
may be several to many Gyr old.  BD+20~307 affords astronomy a rare peek 
into a 
mature planetary system in orbit around a close binary star (because such 
systems are not amenable to study by the precision radial velocity 
technique).

\end{abstract}


\section{INTRODUCTION}

Hundreds of main sequence stars are known with orbiting dusty debris 
disks.  The dust in most of these systems is cold and appears to be 
analogous to dust in the Sun's Kuiper Belt.  Only a small percentage of 
debris 
disks contain warm dust analogous to that in the Sun's asteroid belt and 
zodiacal cloud (e.g., Su et al. 2006; Rhee et al 2008;  Trilling et 
al. 2008; Smith et al 2008).  
For stars that are at least 100 Myr old, the warm dust luminosity 
(L$_{IR}$/L$_{bol}$) is typically a few times 10$^{-4}$ or less.  Smith et 
al (2008) discuss and review the significance of this warm, mid-infrared, 
emission in the context of the Sun's zodiacal dust cloud.  They note that 
{\it Spitzer Infrared Observatory} photometry (Hines et al 2006, Bryden et 
al 2006) indicates that warm dust is detected at only 2$\pm$2\% of the 
observed systems.  A more recent, more comprehensive, {\it Spitzer} survey 
(Trilling et al. 2008) yielded a similar small percentage of warm 
excess emission stars in a sample of 213 F- and G-type stars. And a 
comparable small percentage also applies to warm excess 
A-type stars of age $>$400 Myr (Su et al. 2006) which, as will be seen 
below, is the age range of interest in the present paper. In all of these 
surveys and in other published {\it Spitzer} surveys, the typical 
mid-infrared 
luminosity is sufficiently small that the warm dust could be produced by 
collisions of asteroids; for example, in a massive asteroid belt perturbed 
by the gravitational field of a nearby giant planet.

In contrast, Rhee et al (2008) discuss the dustiest main sequence stars 
currently known: Pleiad HD23514 and, dustiest of all, the G0 field 
dwarf BD+20~307 (HIP8920; Song et al. 2005, Weinberger 2008, Weinberger et 
al. 2008).  The dust luminosity at these two stars is orders of magnitude 
larger than at other mid-infrared excess main sequence stars and  
collisions of asteroids seem quite inadequate to explain the data.  A 
plausible model to account for so much dust so close to 
these stars involves a fairly recent collision of two planets or planetary 
embryos in the terrestrial planet zone (Rhee et al. 2008).
 
BD+20~307 was originally presumed to be a single star, and
a measurement of its (presumably youthful) age would have constrained the 
duration
of the era during which major collisional events occur
in youthful planetary systems.  Using photospheric lithium content,
Galactic space motion, and an upper limit to the X-ray flux
measured with the ROSAT All-Sky Survey, Song et al. (2005)
estimated the age of BD+20~307 to be $\sim$300 Myr.  To refine this
estimate, we obtained X-ray measurements with the {\it Chandra X-ray 
Observatory} and hundreds of photometric optical 
observations to measure the star's rotation period and age, employing the 
technique of gyrochronology (Barnes 2007).  While we were carrying out 
these observations, Weinberger (2008) obtained the surprising result that 
BD+20~307 is a short period ($\sim$3.5 day) spectroscopic binary of two 
nearly identical stars.  The reason this was a surprise is that the Song 
et al. (2005) echelle spectrum showed no evidence of two stars; by a low 
probability ($<$10\%) quirk of fate, their 2004 epoch spectrum was obtained 
at a time when the two stars had nearly the same radial velocities.

While {\it Spitzer} surveys for debris disks around stars have often 
explicitly omitted binaries, Trilling et al. (2007) found small quantities 
of cool dust around a substantial fraction of binary stars.  However, by 
analogy with the discussion of single stars in the first paragraph above, 
dust around such binary stars can readily be accounted for by collisions 
of Kuiper Belt objects.  

Upon learning of the Weinberger (2008) result, we initiated a program to 
monitor the radial velocity of BD+20~307.  Our various observations are 
described in the next Section.  This is followed by a discussion of the 
age and some other aspects of this exceptionally dusty binary star, which
points to the existence of rocky planets in the terrestrial planet region.

\section{OBSERVATIONS}

\subsection{X-rays}

We observed BD+20~307 with the {\it Chandra X-ray Observatory} Advanced 
CCD Spectrometer Spectroscopic array (ACIS-S; Weisskopf et al. 2002) 
starting on 2007 September 25 at 23:34:20 (UT), for a total of 4919 s.  
We reduced the event lists using the Chandra Interactive Analysis of 
Observations (CIAO) software version 3.4 and the calibration database 
version 3.4.1. We filtered the data using the standard techniques 
described on the web site of the {\it Chandra X-ray Center} (CXC).

We then made images of the events in the full bandpass of
0.5--8.0~keV, a soft 0.5--2.0~keV band to provide sensitivity to
foreground sources, and a hard 4--8 keV band to provide sensitivity to
highly-absorbed sources.  We searched these for point sources using
the wavelet-based algorithm {\tt wavdetect} (Freeman et al. 2002). 
Images were made for two spatial scales: one 1024$\times$1024 pixel image
covering the central portion of the field at the 0.5\arcsec\
resolution of the detector, and one image at 2\arcsec\ resolution that
covered all of the active detectors. We identified a total of 30 sources. 
The majority of these, 26, were detected strongly in the
0.5--2~keV band. Only 4 sources were detected in the full band but not
in the soft band, and no sources were detected exclusively in the hard
band. Three of the brightest sources had counterparts in the 2MASS
catalog, so we used these to register the absolute astrometry for the
observation. We found that the X-ray sources were offset by an
average of $-$0.24$\pm$0.17\arcsec\ in right ascension, and
$+$0.36$\pm$0.17\arcsec\ in declination, and applied this shift to the
locations of the X-ray sources. 

We found an X-ray source coincident with BD+20~307 at 01$^h$ 54$^m$
50\fs37, +21\degr\ 18\arcmin\ 22\farcs3, with an uncertainty of
0.4\arcsec according to Equation 5 of Hong et al. (2005). 
We extracted photometry and light curves for the source using the {\tt
acis\_extract} routine.\footnote{{\tt www.astro.psu.edu/xray/docs/TARA/}} 
Events for the source were
extracted from a circle with radius of 2.5 pixels, which corresponds
to 90\% of the PSF at 1.4 keV. Background was derived from an annular
25-pixel region, excluding a circle of 5 pixels around the point source. 
We found 158$\pm$12 net counts from BD+20~307, all of which arrived in 
the 0.5--2.0 keV band. In total, only 4
sources in the 340 arcmin$^2$ field-of-view had $>$100 net counts, so
we conclude that this X-ray source is BD+20~307.

Next, we binned the events over energy to obtain spectra of the
source and background, and obtained the detector response function and
effective area using standard CIAO tools. We modeled the spectrum as a
thermal plasma\footnote{{\tt http://hea-www.harvard.edu/APEC}} using
XSPEC version 12.2.1 (Arnold 1996). We used the Cash statistic
(Cash 1979) to identify a best fit and 1$\sigma$ uncertainty on the 
model parameters, and found a temperature 
$kT$$=$$0.61^{+0.06}_{-0.02}$~keV
and a flux of $1.1\times10^{-13}$ erg cm$^{-2}$ s$^{-1}$ (0.5--2.0 keV). 
Finally, we tested for variability by comparing the arrival times of the
photons to a uniform distribution using the Kolmogorov-Smirnov statistic,
and found that the data were consistent with a constant flux at the 76\%
confidence level.

\subsection{Optical Photometry}

We used the T12 0.8m Automatic Photometric Telescope (APT) located at 
Fairborn Observatory in Arizona to collect a total of 449 nightly 
brightness measurements of BD+20~307 between 2005 February and 2007 
December.  Such measurements of solar-type stars often reveal brightness 
variations caused by cool, dark photospheric spots as they are carried 
into and out of view by the star's rotation (e.g., Radick et al. 1987).  
Our APTs have proven to be very successful at measuring stellar rotation 
periods, even for stars with low photometric amplitudes of a few 
hundredths of a magnitude or less (e.g., see Henry et al. 1995).

We programmed the T12 APT to observe BD+20~307 differentially with respect 
to three comparison stars.  Its two-channel precision photometer observed 
each star simultaneously in the standard Str\"omgren $b$ and $y$ passbands.  
The three comparison stars, designated A, B, and C, respectively, are 
HD~12354 ($V=6.61$, $B-V=0.34$, F1~IV--V), HD~12846 ($V=6.89$, $B-V=0.66$, 
G2~V), and HD~10088 ($V=7.89, B-V=0.31$, F0~IV), while BD+20~307 ($V=8.98, 
B-V=0.56$, F9~V) is designated as star D.  The observations were reduced 
to form the six differential magnitudes D$-$A, D$-$B, D$-$C, C$-$A, C$-$B, 
and B$-$A.  To increase the precision of our measurements, we combined 
the Str\"omgren $b$ and $y$ differential magnitudes into a single $(b+y)/2$ 
passband for each of the six differential magnitudes.  For additional 
information on the telescope, photometer, observing procedures, data 
reduction techniques, and photometric precision, see Henry (1999) and 
Eaton et al. (2003).

The individual reduced and standardized $(b+y)/2$ differential magnitudes 
of BD+20~307 are listed in Table~1, where they will be available for 
future studies such as long-term spot evolution or for comparison with other 
activity indicators in BD+20~307.  To improve our measurement precision 
still further, we averaged the three D$-$A, D$-$B, and D$-$C differential 
magnitudes of BD+20~307 into a single value representing the difference in 
brightness between BD+20~307 and the mean of the three comparison stars 
($D-(A+B+C)/3$), which we refer to as the ensemble average.  The standard 
deviations of the three comparison star differentials C$-$A, C$-$B, and 
B$-$A from Table~1 all lie between 0.001 and 0.002 mag, which is typical of 
the measurement precision with the T12 APT. Therefore, all three 
comparison stars are constant from night to night to the limit of our 
precision.

The 449 ensemble differential magnitudes of BD+20~307 in our $(b+y)/2$ 
passband are plotted in the top panel of Figure~1, where they fall into 
four separate observing seasons.  No significant periodicity was found when 
analyzing the observing seasons separately or the data set as a whole.  
Therefore, any photospheric starspots on either component of BD+20~307 must 
have lifetimes of only a few weeks at most.  Indeed, a small portion of the 
third observing season, marked by the pair of vertical lines in the top 
panel of Figure~1 and shown with an expanded abscissa in the second panel, 
does appear to reveal a few cycles of low-amplitude but coherent brightness 
variability.  The brightest and faintest observations of each cycle are 
connected with straight line segments to guide the eye; the interpolation 
of approximately one and a half missing cycles is shown by the dotted line 
segments.

The third panel of Figure~1 shows the power spectrum of the data from 
the second panel, computed with the method of Van{\'{\i}}{\v c}ek (1971).  
This 
period-finding technique uses least-squares to fit the data with sine 
curves over a range of trial periods and measures the resulting reduction 
in the variance of the data for each trial frequency.  The strongest 
peak occurs at a frequency of $0.2833 \pm 0.0020$ day$^{-1}$, corresponding 
to a period of $3.530 \pm 0.025$ days.  The lower peak is a one-day alias 
of this period.  Analysis of small sections of the fourth season lightcurve 
provides support for the 3.53-day period, yielding less precise period
values between 3.28 and 3.44 days.

The bottom panel of Figure~1 shows the data from the second panel phased 
with the 3.53-day photometric period and a time of brightness minimum of 
JD $2,454,003.18 \pm 0.04$ days, computed from a least-squares sine fit 
to the phase curve.  The sine fit gives a peak-to-peak amplitude of
$0.0066 \pm 0.0005$ mag and an rms to the fit of 0.0014 mag, which 
approximates the typical precision of our photometric measurements.  

\subsection{Optical Spectroscopy}

From 2008 January through March we obtained 28 spectrograms of 
BD$+$20~307 with the Tennessee State University 2 m automatic 
spectroscopic telescope (AST), fiber-fed echelle spectrograph, and a 2048 
x 4096 SITe ST-002A CCD.  The echelle spectrograms have 21 orders, 
covering the wavelength range 4920--7100~\AA\ with an average resolution 
of 0.17~\AA.  The typical signal-to-noise ratio of these observations is 
$\sim$10.  Eaton \& Williamson (2004, 2007) have given a more extensive 
description of the telescope and spectrograph, which are operated at 
Fairborn Observatory.

For the AST spectra, lines in approximately 170 regions, centered on the 
rest wavelengths (Moore et al. 1966) of relatively strong, mostly Fe~I 
lines that were not extensively blended with other nearby strong 
features, were measured.  Lines at the ends of each echelle order were 
excluded because of their lower signal-to-noise ratios.  The wavelength 
scale of each spectrum was initially determined from Th-Ar comparison 
spectra obtained at the beginning and end of the night, and that scale 
was refined with the use of the telluric O$_2$ lines near 6900\AA, which 
are in each stellar spectrum.  A Gaussian function was fitted to the 
profile of each component. If lines of the two components were blended, a 
double Gaussian was used to fit the combined profile.  The difference 
between the observed wavelength and that given in the solar line list of 
Moore et al (1966) was used to compute the radial velocity, and a 
heliocentric correction was applied.  The final mean velocity for each 
component is given in Table~2.  Our unpublished velocities of several IAU 
standard solar-type stars indicate that the Fairborn Observatory 
velocities have a small zero-point offset of $-$0.3 km~s$^{-1}$ relative 
to the velocities of Scarfe et al. (1990).

There is no confusion concerning component identification because the 
line strengths are different enough, as discussed in the next paragraph, 
so that the primary and secondary can be identified visually. With our 
photometric period adopted and 0.3 km~s$^{-1}$ added to each Fairborn 
Observatory velocity, initial orbital elements for the primary were 
computed with BISP (Wolfe et al. 1967), a program that implements a 
slightly modified version of the Wilsing-Russell method.  The orbit was 
then refined with SB1 (Barker et al. 1967), a program that uses 
differential corrections.  An orbit for the secondary velocities was also 
computed. Zero weight was assigned to the velocities of the nine 
observations that had blended components with velocity separations less 
than 30 km~s$^{-1}$. Based on the variances of the two solutions, weights 
of 1.0 and 0.9 were assigned to the primary and secondary velocities, 
respectively, of the 19 other spectrograms.  An orbital solution of both 
components together was obtained with SB2, a modified version of SB1. 
Because the eccentricity of this solution is extremely small, 0.004 $\pm$ 
0.004, a circular orbit was computed with SB2C (D. Barlow 1998, private 
communication), which also uses differential corrections to determine the 
elements.  For this solution the weights of the velocities were the same 
as those previously assigned. The tests of Lucy \& Sweeney (1971) 
indicate that the circular orbit solution is to be prefered.  The orbital 
phases of the observations and velocity residuals to the final computed 
curves are given in Table~2, and the orbital elements and related 
parameters are listed in Table~3.  For a circular orbit the element {\it 
T}, a time of periastron passage, is undefined.  So, as recommended by 
Batten et al. (1989), {\it T$_0$}, a time of maximum velocity for the 
primary, is given instead.  In Figure~2 the observed velocities are 
compared with the computed velocity curves.  Zero phase is a time of 
maximum velocity of the primary.

The components are similar in spectral type, as indicated by the best fit 
effective temperatures and gravities that were determined by Weinberger 
(2008).  Thus, for the two components the iron lines, which dominate the 
spectrum, and those of other elements except for lithium, which is 
discussed in \S3, are similar in strength.  To determine the magnitude 
difference between the two components, we followed the spectroscopic 
method of Petrie (1939).  For spectra with well separated components we 
measured the ratio of the equivalent widths of over 100 Fe~I primary and 
secondary lines.  The average of the equivalent width ratio of the 
secondary to primary components is 0.78 $\pm$ 0.01 and corresponds to the 
luminosity ratio of the stars.  This is then converted into a very modest 
magnitude difference of 0.27.  Given the wavelength range of our spectra, 
we adopt this value as the $V$ magnitude difference.  Combined with the 
total $V$ mag of 8.98 listed by {\it Hipparcos}, this results in $V_A$ = 
9.61 mag and $V_B$ = 9.88.

Weinberger (2008) compared synthetic spectra, computed with solar 
abundances, to a spectrum of BD$+$20~307 with well separated components 
and determined the effective temperatures of the two stars.  The two 
components differed by just 250 K, and the average of the two effective 
temperatures is 6375 K, which corresponds to a $B-V$ color of 0.48 
(Flower 1996).  However, the observed combined $B-V$ color of the system 
is 0.56 (Perryman et al. 1997).  This suggests either that the star has 
significant extinction in the visual region or the components are metal 
poor.  The first possibility appears to be unlikely.  At a distance of 
only 102 pc (van Leeuwen 2007), chances are small that sufficient 
interstellar matter lies between the Sun and BD$+$20~307 to produce a 
color excess of 0.08 mag (Welsh et al. 1998; Lallement et al. 2003; Cox 
2000; B. Welsh 2008, personal communication).  Also, little, if 
any, circumstellar extinction from the collision that produced the large 
infrared excess is expected.  Collisions among the dust particles rapidly 
damp out the components of motion not in the orbital plane.  For 
BD$+$20~307 the plane containing the collision material is presumably the 
same as that of the binary orbit, which has an inclination of about 
31\arcdeg\ and so is much closer to face-on ($i$ = 0\arcdeg) than 
edge-on.  In addition, the spectral energy distribution of BD$+$20~307 is 
such that very little (hot) dust exists within a few tenths of an AU of 
the central stars (Song et al. 2005, Weinberger et al. 2008).  Thus, the 
picture of this system is of the close binary components shining out 
through the central hole in the dust ring distribution.

The metal poor alternative appears to be a more attractive explanation of 
the apparently discrepant B-V indicated in the previous paragraph.  
Synthetic spectra with solar abundances require a higher effective 
temperature then do metal poor stars to produce a given (weak) line 
strength. To investigate whether the system is metal poor, we estimated 
the iron abundance for the components of BD$+$20~307. We compared the 
combined average equivalent width of components A and B, determined from 
the same set of iron lines as those used for the luminosity ratio, with 
the average iron equivalent widths of several single dwarf stars, 
including $\chi$~Her, $\beta$~Com, and HR~7560, which have similar $B-V$ 
values and known iron abundances.  The spectra of these comparison stars 
were obtained with the same telescope, spectrograph, and detector as 
those of BD$+$20~307.  The average equivalent widths of HR~7560 and 
$\beta$~Com, which have essentially solar iron abundances (Taylor 2005), 
are 30\% larger, but the average equivalent width of $\chi$~Her, which 
has [Fe/H] = $-$0.43 (Taylor 2003), is identical to that of BD~$+$20~307.  
Thus, we conclude that BD~$+$20~307 has a similar iron abundance and is 
metal poor compared to the Sun.  Given that dust particles are made of 
refractory metals, it does seem remarkable that the dustiest main 
sequence star presently known to astronomy appears to be of low 
metallicity.

\section{DISCUSSION}

The initial motivation for the research reported in this paper was 
determination of the age of BD+20~307 and the implications of its very 
dusty circumstellar disk for the formation of terrestrial planets around 
sun-like stars.  As such, we 
initially obtained the X-ray and optical photometry data described in 
Sections 2.1 and 2.2.  However, the discovery that BD+20~307 is a close 
binary star (SB2; Weinberger 2008) greatly altered our interpretation 
of these data and motivated the spectroscopic observations discussed in 
Section 2.3.  In the present Section we consider some properties of 
BD+20~307 derived from the data and their relatationship to its age.

Assuming BD+20~307 is a single star, from a fit to its spectral energy 
distribution, Song et al (2005) derived a radius = 1.25 R$_\sun$ based on 
the Hipparcos parallax of 10.88 mas, and 
temperature 6000~K.  Because of its low metallicity ([Fe/H] = $-$0.43, 
Section 2.3) the Hipparcos B$-$V = 0.56 should be corrected to a solar 
metallicity B$-$V index according to an expression given in Gray (1994); 
the result is B$-$V = 0.607, corresponding to a temperature of $\sim$5900 
K.  With this effective temperature, the monochromatic Hipparcos V mag 
(= 8.98), and the Hipparcos parallax, we derive a single star radius of 1.3 
R$_\sun$, in good 
agreement with the results of Song et al.  Van Leeuwen (2007) revised 
the Hipparcos parallax to 9.82 mas, which corresponds to a single star 
radius of 1.44 R$_\sun$.  With the simplifying 
assumption that the two stars in BD+20~307 are of equal magnitude, then 
their representative radii would each be 1 R$_\sun$.  (More precise 
values for the radii, appropriate for the relative magnitudes given  
in \S2.3, can easily be derived.) Conversion of T and R to 
stellar mass depends on the uncertain stellar age (see below).  If, the 
mass of the primary is 1 M$_\sun$, then the orbital inclination $i = 
31\arcdeg$.

Our orbital period for a circular orbit (Table~3) agrees well with the
period of 3.448 days derived by Weinberger (2008) for an assumed circular
orbit.  However, her systemic radial velocity, $\gamma$ = $-$8.4
km~s$^{-1}$, does not agree within the respective errors with our $-$12.43
km~s$^{-1}$.  From Weinberger's discussion, the quoted error on her value
for $\gamma$ would be about 0.6 km~s$^{-1}$.  Then, the $>$6 $\sigma$
discrepancy in radial velocities must be attributed to either an
underestimate of measurement error or a variable radial velocity or both.  
To investigate the possibility of a variable radial velocity, we divided
our Table 3 data set into two halves each 22 days long.  The result was
$\gamma$ = $-$12.4 km~s$^{-1}$ for each half, with estimated uncertainties
of 0.1 and 0.2 km~s$^{-1}$, for the first and second half, respectively.  
Thus, if the systemic radial velocity of BD+20~307 is variable, then it
changed by about 4 km~s$^{-1}$ over the 117 day interval between the
2007 October observations of Weinberger and the first half of our data set
but then essentially did not change over the 22 days that separated the
two halves of our data.  While this pattern seems quite peculiar, it
cannot be ruled out a priori.  We note that in September 2004, Song et al.
(2005) measured a systemic radial velocity of $-$11$\pm$1 km~s$^{-1}$.  
Tokovinin et al. (2006) investigated spectroscopic binaries for the 
presence of tertiary companions.  For a system with orbital period of 3.4 
days, the 
probability is $\sim$70\% that a third star will also be present. The 
period distribution is such that, depending on mass 
ratios, if a third star at BD+20~307 exists, then it might be detected 
either through adaptive optics imaging or monitoring of the 
systemic radial velocity.

Our observed orbital period of 3.42 days is in reasonable agreement with
our (less precisely) determined photometric period of 3.53 days, as would
be expected for a system with synchronized rotation and revolution.  As
noted by Weinberger (2008 and references therein), for solar mass stars
and an orbital period of 3.4 days, this synchronization would occur in
$\sim$100 Myr which is likely to be substantially less than the age of
BD+20~307 (see below).  Quasi-sinusoidal lightcurves are typical of
spot-induced variability in magnetically active stars (e.g., Henry et al.
1995; Henry \& Winn 2008). The low photometric amplitude implies that the
difference in spot filling factor over the observable hemisphere during a
full rotation period is less than 1\%.  The moderately small inclination
of the rotation (orbital) axis, $\sim$30$\arcdeg$, would be consistent 
with the low photometric amplitude of the lightcurve.  Mid-infrared
interferometric observations, for example with the Keck Nuller
Interferometer, might confirm or deny whether the debris
disk around BD+20~307 lies in the orbital plane of the binary.

For main sequence stars with outer convection zones, both rotation rate and 
associated stellar activity (e.g., X-ray flux, optical emission lines) can 
be used to derive stellar ages (see, for example, Gaidos 1998; Zuckerman \& 
Song 2004; Barnes 2007).   While the rotation rate of BD+20~307 is 
controlled by its orbital period and not by its age, it is nonetheless 
worthwhile to characterize the binary's X-ray luminosity.

With the effective single-star radius and temperature calculated in the 
second paragraph of the present Section and the Chandra-measured X-ray 
flux in the $0.5-2.0$ keV band quoted in Section 2.1, we calculate a 
fractional X-ray luminosity (L$_x$/L$_{bol}$) of $1.6\times10^{-5}$.  This 
may be compared to the upper limit to the ROSAT measured X-ray flux over a 
similar energy band quoted by Song et al (2005), i.e., $1.26\times10^{-5}$. 
Thus, it appears that the (presumably somewhat variable) X-ray flux from 
BD+20~307 was just below the ROSAT detection limit.  The high X-ray flux 
and the rapid rotation period of 3.4 days are consistent with the fact that 
this system is a close, synchronized binary.

In the absence of rotation and activity age indicators, one can instead
turn to lithium abundance, Galactic space motion (UVW), metallicity, and
placement on evolutionary tracks on an HR diagram.  With the lithium
absorption feature equivalent width (EW) given by Weinberger (2008) and
our observed ratio of the two stellar continua (= 0.78) at 6708 \AA\, we
derive an EW of 62 m\AA\ for the primary.  This would be consistent with
an age perhaps as young as stars in the UMa or Hyades groups (i.e.,
$\sim$500 Myr, e.g., Zuckerman et al. 2006), but also consistent with
early G-type stars in the 2 Gyr old NGC 752, IC 4651, and NGC 3680 open
clusters (Sestito et al. 2004).  Indeed, this EW is even similar to that
of early G-type stars in the 6$-$8 Gyr old NGC 188 (Randich et al. 2003).
Then there is the secondary to consider, for which the 3 $\sigma$ EW is
$<$14 m\AA\ (Weinberger 2008).  Such small EW have been measured for some
early G-type stars as old as $\sim$5 Gyr, for example M67 (Sestito \&
Randich (2005) and the Sun, and perhaps occasionally also as young as 2
Gyr (NGC 752, Sestito et al. 2004).  Thus, one can conclude that the
lithium lines in the two stars are probably most consistent with an age of
a few Gyr or more.

With the van Leeuwen (2007) parallax, the Tycho~2 proper motion, and a 
radial velocity of -12.4 km s$^{-1}$, we calculate the Galactic space 
motion of BD+20~307 with respect to the Sun (UVW = -3.3, -24.5, +4.7 km 
s$^{-1}$).  This UVW is more similar to the UVW of moving groups with ages 
$\leq$100 Myr than it is to the UVW of older moving groups such as Ursa 
Majoris and the Hyades (see, e.g., Table 7 in Zuckerman \& Song 2004).  
While relatively few stars of age several Gyr or more have such small 
absolute values of all three space motion components, it is of course 
possible that BD+20~307 is one such star.

The low metallicity of BD+20~307 suggests an old age.  Even the 5 open 
clusters with ages between 2 and 8 Gyr listed by Sestito \& Randich (2005) 
all have metallicities not much different than solar.\footnote{Lambda Boo 
stars have low abundances of refractory elements and dusty/gaseous disks 
and are believed to be young, age $\leq$few 100 Myr.  Their spectral 
types, late-B to early-F, are substantially earlier than that of 
BD+20~307.  Thus, BD+20~307 is very unlikely to be a (young) $\lambda$ 
Boo type star.}  With the 
evolutionary tracks presented by Nordstrom et al. (2004; Fig. 12, [Fe/H = 
-0.50] panel), and the van Leeuwen (2007) parallax, BD+20~307 looks to be 
many Gyr 
old.  However, the parallax error is sufficiently large, 13\%, that, 
within the uncertainties, a much 
younger age is consistent with these tracks.  A very accurate 
measurement of the parallax of BD+20~307 would be worthwhile.

In summary, various methods taken together seem to imply a BD+20~307 age
of least a few Gyr, and possibly considerably more.  It is unlikely that
the extraordinary quantity of warm dust at BD+20~307 has any direct
relationship with the origin of terrestrial planets.

\section{CONCLUSIONS}

Until new techniques are available to image or otherwise detect planets 
with masses and semi-major axes similar to those of the terrestrial planets 
of our solar system, we must rely on more indirect observations and models.  
One promising indirect approach is detection of dusty debris associated 
with terrestrial planet formation (e.g., Gorlova et al. 2007; Currie et al. 
2007; Rhee et al. 2008; Lisse et al. 2008).  To delineate the full extent 
of the era of terrestrial planet formation it is necessary to identify main 
sequence stars with copious quantities of warm dust grains and ages of tens 
to many hundreds of millions of years.  Thus, the goal of our research 
project was to establish the age of BD+20~307 which, along with HD 23514 
(an F6 member of the Pleiades), are the only two stars known with age 
$\geq$100 Myr and strong indications of the existence of orbiting 
terrestrial planets (Rhee et al. 2008).
 
Based on the findings of our study, it now appears likely that BD+20~307
is an old binary star and its copious quantity of warm orbiting dust has
no direct relationship with the era of planet formation.  Rhee et al.
(2008)  present a moderately detailed discussion of the origin and fate of
the dust at HD 23514 and BD+20~307.  Given the large quantity of dust at
BD+20~307, its short lifetime in orbit, and its concentration to a narrow
range of warm temperatures (Weinberger et al. 2008), it is hard to escape
the conclusion that something with the mass of a terrestrial planet was
involved in a catastrophic collision, independent of whether that
collision took place between two isolated objects or inside some sort of
massive asteroid belt.  A remarkable aspect of BD+20~307 is that this
collision apparently has taken place so late in its history.  Perhaps, as
suggested by the discussion of possible systemic radial velocity
variations in Section 3, there is an unseen stellar or brown dwarf object
present whose gravitational influence recently destabilized the orbits of
rocky objects of planetary mass.

In any event, whatever its age, if the massive amounts of dust at 
BD+20~307 do point toward the presence of terrestrial planets, then this 
represents the first known example of planets of any mass in orbit 
around close binary stars -- because the precision radial velocity 
technique is not sensitive to planets in such systems. 

Because of its only modest sensitivity, IRAS was quite limited in the distance 
out to which it could detect warm dust at low luminosity, main sequence, stars 
of K- and M-type.  Should they exist, such very dusty, low luminosity, stars 
will be detectable by the Wide-Field Infrared Survey Explorer (WISE) 
mission.  A substantial sample of very dusty stars of
known age could delineate the era of terrestrial planet formation while
perhaps revealing the existence of additional remarkable stars like 
BD+20~307.

\acknowledgments
We thank Inseok Song for a helpful conversation and Alycia Weinberger for 
calling the binary nature of BD+20~307 to our attention before 
publication.  We thank the second referee for constructive comments.
This research was supported in part by NASA through a Chandra Observatory 
award to UCLA.  Automated Astronomy at Tennessee State University is 
supported
by NASA, NSF, Tennessee State University, and the State of
Tennessee through its Centers of Excellence program.

{}

\clearpage
\begin{deluxetable}{lrrrrrr}
\tabletypesize{\normalsize}
\tablecaption{Individual Photometric Observations of BD+20~307}
\tablewidth{0pt}
\tablehead{
\colhead{Hel. Julian Date} & \colhead{D$-$A} & \colhead{D$-$B} & \colhead{D$-$C} & \colhead{C$-$A} & \colhead {C$-$B} & \colhead {B$-$A} \\
\colhead{(HJD $-$ 2,400,000)} & \colhead{(mag)} & \colhead{(mag)} & \colhead{(mag)} & \colhead{(mag)} & \colhead{(mag)} & \colhead{(mag)}
}
\startdata
53,402.5866 & 2.4388 & 2.0525 & 1.1812 & 1.2576 & 0.8712 & 0.3864 \\
53,402.5922 & 2.4401 & 2.0518 & 1.1816 & 1.2585 & 0.8703 & 0.3881 \\
53,404.5932 & 2.4366 & 2.0495 & 1.1767 & 1.2598 & 0.8728 & 0.3871 \\
53,405.5882 & 2.4385 & 2.0507 & 1.1799 & 1.2586 & 0.8708 & 0.3878 \\
53,405.5939 & 2.4363 & 2.0516 & 1.1766 & 1.2598 & 0.8750 & 0.3848 \\
\enddata

\tablecomments{Differential magnitudes were measured in the 
Str\"omgren $(b+y)/2$ passband.  This table is available in its
entirety in machine-readable form in the online journal.  A portion 
is shown here for guidance regarding its form and content.}
\end{deluxetable}

\clearpage
\begin{deluxetable}{lcrrrr}
\tablenum{2}
\tablewidth{0pt}
\tablecaption{RADIAL VELOCITIES OF BD +20 307}
\tablehead{ \colhead{Hel. Julian Date} & \colhead {} & \colhead{$V_A$} &
\colhead{($O-C$)$_A$} & \colhead{$V_B$} & \colhead{($O-C$)$_B$}  \\
\colhead{(HJD $-$ 2,400,000)} & \colhead {Phase} & 
\colhead{(km~s$^{-1}$)} &
\colhead{(km~s$^{-1}$)} &\colhead{(km~s$^{-1}$)} &\colhead{(km~s$^{-1}$)}
}
\startdata
 54,495.585 & 0.842 &    10.8 & $-$0.4 & $-$37.1 &    0.7 \\
 54,497.586 & 0.427 & $-$51.4 & $-$0.2 &    29.8 &    0.6 \\
 54,498.586\tablenotemark{a} & 0.720 & $-$23.4 & $-$2.8 & $-$4.8 & $-$1.2 
\\ 
 54,498.690\tablenotemark{a} & 0.750 & $-$12.9 & $-$0.5 & $-$12.9 & 
$-$0.5 \\ 
 54,499.586 & 0.012 &    30.6 & $-$0.1 & $-$58.7 &    0.0 \\ 
 54,499.690 & 0.042 &    28.4 & $-$0.9 & $-$57.2 &    0.0 \\ 
 54,502.690 & 0.919 &    26.1 &    0.7 & $-$52.6 &    0.4 \\ 
 54,505.630\tablenotemark{a} & 0.779 & $-$3.4 &    1.2 & $-$20.0 & 0.9 \\ 
 54,505.679\tablenotemark{a} & 0.793 & $-$1.6 & $-$0.8 & $-$24.6 & 0.3 \\ 
 54,506.630 & 0.071 &    26.2 & $-$0.3 & $-$54.3 &  $-$0.1 \\ 
 54,506.679 & 0.086 &    24.9 &    0.3 & $-$52.2 &    0.0  \\ 
 54,507.590 & 0.352 & $-$38.2 &    0.1 &    14.7 &  $-$0.7 \\
 54,508.590 & 0.645 & $-$38.0 &    1.0 &    16.6 &    0.5 \\ 
 54,510.591\tablenotemark{a} & 0.230 &  $-$3.3 &   3.6 & $-$22.2 & $-$3.9 
\\ 
 54,514.593 & 0.400 & $-$47.7 & $-$0.4 &    25.3 &    0.2  \\ 
 54,516.594 & 0.985 &    30.1 & $-$0.5 & $-$59.4 &  $-$0.8 \\            
 54,516.658 & 0.004 &    30.7 & $-$0.1 & $-$59.6 &  $-$0.8 \\            
 54,517.594\tablenotemark{a} & 0.277 & $-$20.4 & $-$0.6 &  $-$5.7 &  
$-$1.2 \\ 
 54,517.658\tablenotemark{a} & 0.296 & $-$23.4 &    1.3 &     1.4 &    
0.6 \\ 
 54,522.597\tablenotemark{a} & 0.740 & $-$12.4 &    2.7 & $-$12.4 &  
$-$2.9 \\ 
 54,523.597 & 0.032 &    29.9 &    0.0 & $-$58.1 &  $-$0.2 \\ 
 54,523.648 & 0.047 &    29.8 &    0.9 & $-$55.8 &    1.0 \\ 
 54,524.598 & 0.325 & $-$31.6 &    0.5 &     8.8 &    0.2 \\ 
 54,532.602 & 0.665 & $-$34.2 &    0.1 &    10.7 &  $-$0.4 \\ 
 54,533.604 & 0.958 &    29.9 &    0.6 & $-$56.0 &    1.2 \\ 
 54,534.604\tablenotemark{a} & 0.251 & $-$11.8 &    0.8 & $-$11.8 &    
0.4 \\ 
 54,535.604 & 0.543 & $-$54.9 & $-$0.8 &    31.6 &  $-$0.7 \\ 
 54,540.602 & 0.004 &    29.9 & $-$0.9 & $-$60.0 &  $-$1.2 \\ 
\enddata
\tablenotetext{a}{Spectrum has blended components, velocities given zero 
weight}
\end{deluxetable}

\clearpage
\begin{deluxetable}{lc}
\tablenum{3}
\tablewidth{0pt}
\tablecaption{ORBITAL ELEMENTS AND RELATED PARAMETERS OF BD +20 307}
\tablehead{ \colhead{Parameter} & \colhead {Value } \\
}
\startdata
$P$ (days)             & 3.42015 $\pm$ 0.00067 \\
$T_0$ (HJD)              &  2,454,506.3855 $\pm$ 0.0028 \\
$\gamma$ (km~s$^{-1}$) & $-$12.43  $\pm$ 0.11 \\
$K_A$ (km~s$^{-1}$)      & 43.20  $\pm$ 0.18  \\
$K_B$ (km~s$^{-1}$)      & 46.39  $\pm$ 0.18  \\
$e$                    &   0.0 (adopted)  \\
$M_A$~sin$^3$~$i$ ($M_{\sun}$) & 0.1322  $\pm$ 0.0012 \\
$M_B$~sin$^3$~$i$ ($M_{\sun}$) & 0.1231  $\pm$ 0.0011 \\
$a_A$~sin~$i$ (10$^6$ km)     & 2.0317 $\pm$ 0.0083   \\
$a_B$~sin~$i$ (10$^6$ km)     & 2.1816 $\pm$ 0.0087   \\
Standard error of an observation of unit weight (km~s$^{-1}$) & 0.6 \\
\enddata
\end{deluxetable}

\clearpage
\begin{figure}[p]
\epsscale{0.8}
\plotone{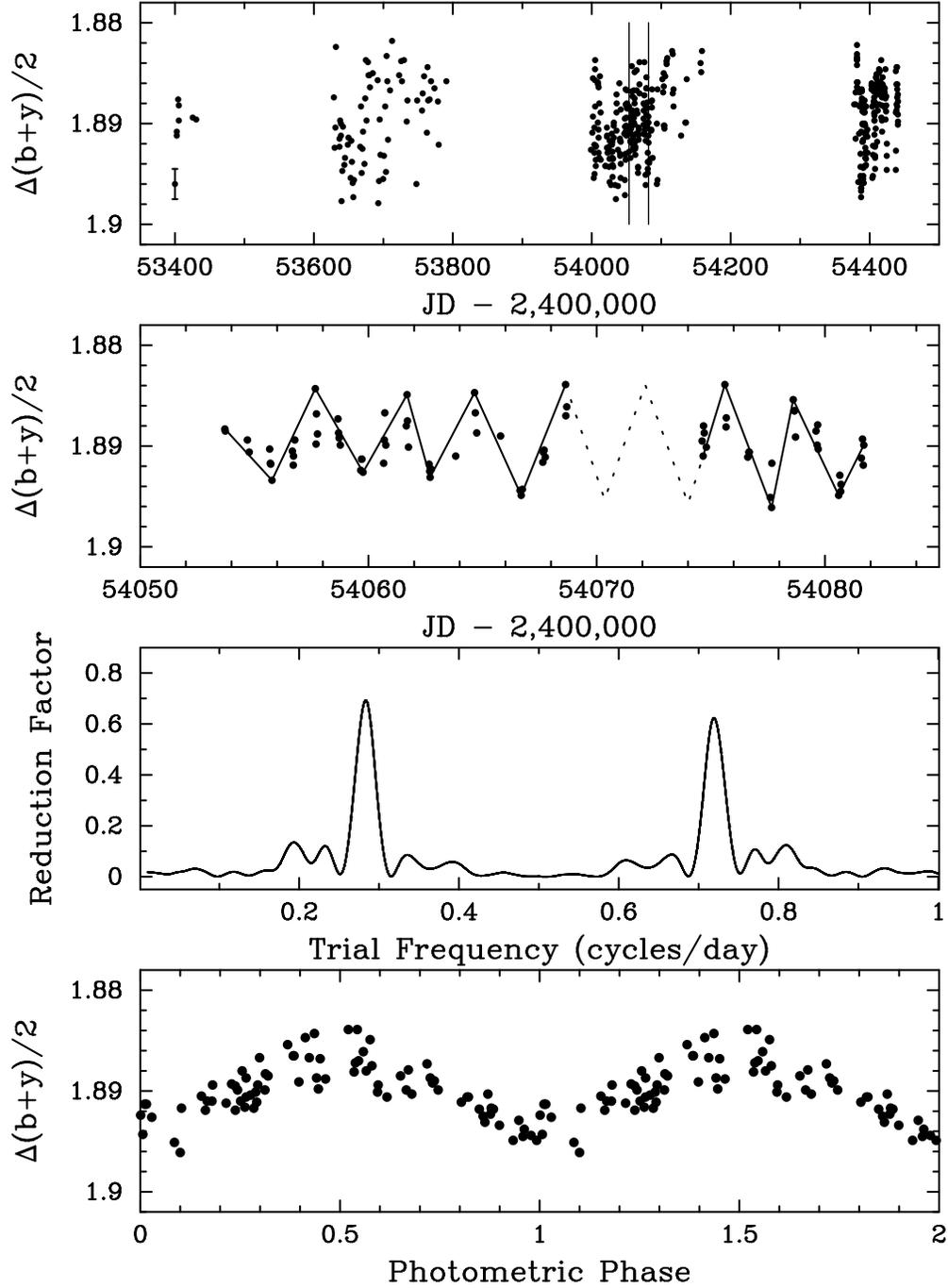}
\caption{$Top~Panel$:  The ensemble photometric observations of BD+20~307 
taken in the Str\"omgren $(b+y)/2$ passband with the T12 0.8m APT at 
Fairborn Observatory over four observing seasons. The error bar in the 
lower left corner represents the typical 1-$\sigma$
uncertainty ($\pm 0.0015$ mag) for a single observation.  
$Second~Panel$:  A 
portion of the third observing season, marked with the vertical lines in 
the top panel and plotted with an expanded absissa, shows low-amplitude 
but coherent variability in BD+20~307.  $Third~Panel$:  Power spectrum of 
the data in panel 2 revealing a best period of 3.530 days.  $Bottom~Panel$:  
Plot of the data in the second panel phased with the 3.530-day period and a 
computed time of minimum.  The amplitude of brightness variation is only 
0.0066 mag.}
\end{figure}

\clearpage
\begin{figure}[t!]
\figurenum{2}
\epsscale{0.8}
\plotone{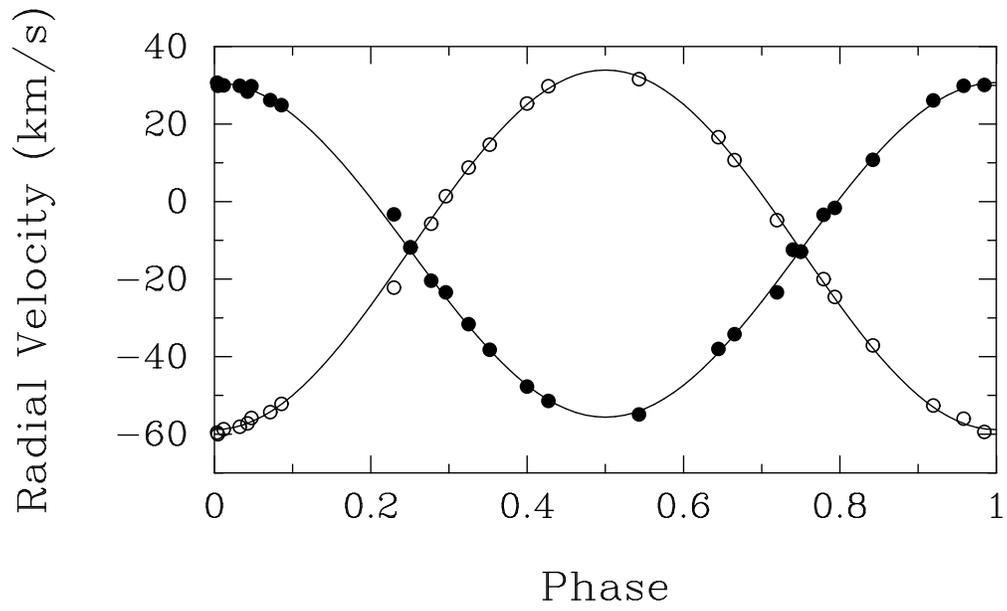}
\figcaption{Radial velocities (dots = component A, open circles =
component B) of BD +20 307 compared with the computed radial 
velocity curves. 
Zero phase is a time of maximum velocity of the primary, component A.
\label{fig2}}
\end{figure}

\end{document}